\documentstyle[prl,preprint,aps,psfig]{revtex}
\begin{document}
\draft

\title{Cooper Instability of Composite Fermions}
\author{Vito W. Scarola, Kwon Park, and J.K. Jain}
\address{Department of Physics, 104 Davey Laboratory,
The Pennsylvania State University, University Park, Pennsylvania 16802}

\date{\today}

\maketitle

{\bf When confined to two dimensions and exposed to a strong 
magnetic field, electrons screen the Coulomb interaction in a 
topological fashion; they capture and even number of quantum vortices and transform into particles called `composite fermions'\cite {Review1,Review2,Jain}.  The fractional quantum Hall effect \cite{Tsui} occurs in such a system when the ratio (or
 `filling factor', $\nu$) of the number of electrons and the degeneracy of 
their spin-split energy states (the Landau levels) takes on particular values.  The Landau level filling $\nu=1/2$ corresponds to a metallic state in which
the composite fermions form a gapless Fermi sea 
\cite {HLR,Willett1,Goldman,Kang}.  
But for $\nu=5/2$, a fractional quantum Hall effect is observed instead
\cite {Willett,Pan};  this unexpected result is the subject of considerable 
debate and controversy \cite {Eisenstein}.  Here we investigate the difference
 between these states by considering the theoretical problem of two composite 
fermions on top of a fully polarized Fermi sea of composite fermions.  We find
that they undergo Cooper pairing to form a $p$-wave bound state at $\nu=5/2$, 
but not at $\nu=1/2$.  In effect, the repulsive Coulomb interaction between electrons is overscreened in the $\nu=5/2$ state by the formation of composite 
fermions, resulting in a weak, attractive interaction.
}

The most important property of composite fermions
is that they do not experience the external magnetic 
field $B$ but rather a drastically reduced magnetic field 
$B^*=B-2p\rho\phi_0.$  Here $\rho$ is the two-dimensional density of fermions, 
$2p$ is the number of vortices, carried by the composite fermion, often intuitively  
envisioned as $2p$ flux quanta, and $\phi_0=h/e$ is a flux quantum.
In effect, each electron absorbs $2p$ flux quanta of the external field to turn 
into a composite fermion.  Electrons confined in two dimensions have unusual 
properties in a magnetic field;
considerable progress towards understanding these  properties has been made by modeling the 
composite fermion (CF) system as an non-interacting gas of composite fermions 
\cite{Review1,Review2}.
In particular, the fractional quantum Hall effect (FQHE) \cite {Tsui}
is a manifestation of the integral quantum Hall effect of composite
fermions \cite{Jain}, and the metallic, non-FQHE state at the half-filled lowest Landau level is a Fermi sea of composite fermions 
\cite {HLR,Willett1,Goldman,Kang}.

The Landau level filling $\nu=5/2=2+1/2$
corresponds to half-filled second LL.
Here, both spin states of the lowest Landau level (LL) are completely occupied, 
contributing 2 to the filling factor.
The fully occupied LL is treated as inert in this work, and the electrons in the partially filled 
second LL are assumed to be fully spin-polarized.  These are valid approximations in  
sufficiently high magnetic fields.
In complete analogy to the half-filled lowest LL, the 
model of non-interacting composite fermions would predict a Fermi sea
of composite fermions at $\nu=5/2$ as well.
However, experiments reveal a FQHE state here \cite{Willett,Pan}.  In fact, 
5/2 is the only even-denominator fraction to be observed in a single-layer system, 
and its physical origin has been a subject of debate and controversy \cite{Eisenstein}.

To understand
the fundamental difference between $\nu=5/2$ and $\nu=1/2$, 
it is necessary to go beyond the model of non-interacting composite fermions.
Our theoretical investigations of the inter-CF interaction
employ the Jain wavefunctions for composite fermions\cite{Review1,Review2,Jain}.  These wavefunctions not only give an accurate quantitative 
account of the inter-CF interaction, but even capture the 
subtle, interaction-driven Wigner and Bloch instabilities of the CF liquid at small 
filling factors \cite {Kamilla,Park};  such instabilities are analogous 
to those believed to occur for the ordinary 
electrons gas (jellium) at low densities \cite {Ceperley}.

\begin{figure}
\centerline{\psfig{figure=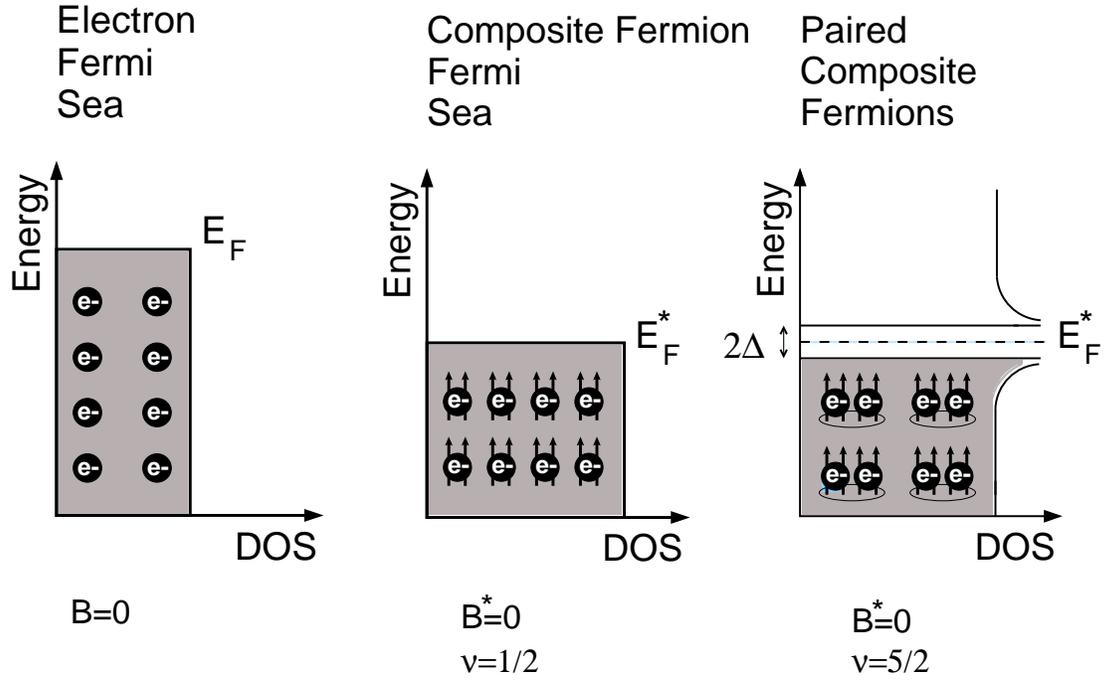,width=16.0cm,angle=-90}}
\caption{Density of States (DOS) for electrons and composite fermions at zero 
effective magnetic flux.  Left panel, electrons at $B=0$.  Centre panel, 
composite fermions at $\nu=1/2$.  Right panel, pairing of composite fermions 
at $\nu=5/2$.  The composite fermions are shown as electrons carrying two flux
quanta.
\label{fig1}}
\end{figure}

Central to this work is the following question, in analogous to the Cooper problem for ordinary
superconductivity: if we begin by assuming a Fermi sea of composite fermions both 
at $\nu=1/2$ and $\nu=5/2$ and add two composite fermions at the Fermi surface,
will they form a bound state?  
We find that the CF-Fermi sea is unstable to pairing of composite fermions at 
$\nu=5/2$ but not at $\nu=1/2$, as shown schematically in Fig.~(\ref{fig1}).
We stress that, in contrast to the Bardeen-Cooper-Schrieffer (BCS) theory, we do not assume any attractive 
interaction, phonon-mediated or otherwise; the only interaction in the problem 
is the repulsive Coulomb interaction between electrons.
However, the Coulomb interaction translates into a weak attractive 
interaction between composite fermions at $\nu=5/2$.

We work in the spherical geometry \cite {Haldane,Yang}, in which N electrons 
are considered to move on the surface of a sphere under the presence of a radial
magnetic field produced by a magnetic monopole of strength $Q$ at the centre.  
The flux through the surface of the
sphere is $2Q\phi_{0}$, where $2Q$ is an integer, according to Dirac's quantization 
condition.  The composite fermion theory maps the problem of interacting
electrons at $Q$ to that of composite fermions at $Q^*=Q-N+1$ (assuming 
here and below composite fermions of vorticity $2p=2$).
The Jain wavefunctions for interacting electrons at $Q$ are given by  
$\Psi_Q={\cal P}_{LLL}\Phi_1^2\Phi_{Q^*},$
where $\Phi_{Q^*}$ are wavefunctions of non-interacting electrons at ${Q^*}$,
and $\Phi_1$ is the wavefunction of the fully occupied lowest Landau level
at $Q_1=(N-1)/2$,
and ${\cal P}_{LLL}$ is the lowest LL projection operator.
Due to the rotational symmetry, the total orbital angular momentum $L$ is a good
quantum number, preserved in going from $\Phi_{Q^*}$ to $\Psi_Q$. 

We are interested in composite fermions in a vanishing effective magnetic field, that is, when $Q^*=0$, which is obtained at $Q=N-1$. 
Here, for $N=n^2$, the ground state 
$\Psi_Q$ ($\Phi_{Q^*}$) has uniform density ($L=0$), as it contains $n$ 
filled shells of composite fermions (electrons).  We approach the CF Fermi sea as the $N\rightarrow \infty$ limit of the filled shell states.  
The systems with a CF pair occur at $Q^*=0$ for $N=n^2+2$, corresponding to 
situation when two CF particles are added to the $(n+1)$st shell.
The individual angular momenta of the additional particles are $l=n$, implying that there
is one multiplet at each total angular momentum $L=1,3,...,L_{max}=2n-1$ with  a degeneracy
of $2L+1$.  The wavefunction of a CF pair at $B^*=0$ for a given $L$ 
is $\Psi_L^{CF-pair}={\cal P}_{LLL}\Phi_1^2\Phi_L^{el-pair}$, where $\Phi_L^{el-pair}$ is 
the wavefunction of an electron pair at $B=0$.  
We will also consider a pair of CF, corresponding to two holes in the 
$n$th shell at $Q^*=0$.  $\Psi_L^{CF-pair}$ contains no adjustable parameter.

For $\Phi_L^{el-pair}$, the Coulomb energy of the electron pair, which is proportional 
to the average inverse distance between the two electrons in the otherwise empty 
shell,  decreases with $L$, in accordance with Hund's rule of atomic 
physics \cite{Quinn}.
This implies that the the  
smallest $L$ ($L=1$ for fully polarized particles) corresponds to the smallest
 distance between two electrons of a pair.  By analogy, the size of the CF pair 
in $\Psi_L^{CF-pair}$ also increases with $L$.  We have confirmed this by monitoring the
influence of an additional short-range interaction to the pair energy.

As it stands, $\Psi_L^{CF-pair}$ is written in the lowest Landau level, that is, 
for $\nu=1/2$. In order to treat $\nu=5/2$, we use the method of
Park {\em et al.} \cite{PMBJ} to map the problem of the Coulomb interaction in the second
($s=1$) Landau level
into that of an effective interaction $V^{eff}(r)$ in the lowest ($s=0$) Landau level.  $V^{eff}(r)$ is 
chosen so that the two interactions have the same Haldane 
pseudopotentials\cite{Haldane} that is, $V_{1,m}=V^{eff}_{0,m}$.
The parameters $V_{s,m}$ are the interaction energies of two particles in $s$th LL in
relative angular momentum $m$ state (in the planar geometry) and completely specify the 
interparticle interaction in the $s$th Landau level.

\begin{figure}
\centerline{\psfig{figure=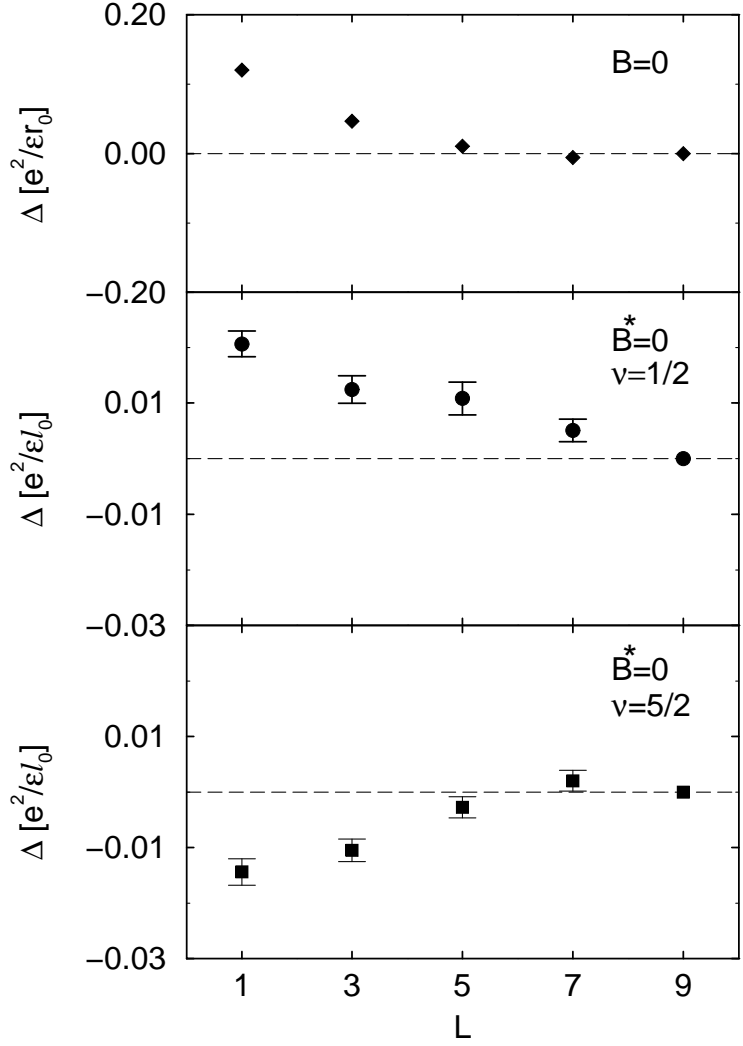,width=12.0cm,angle=0}}
\caption{The interaction energy of pairs of electrons and pairs of 
composite fermions at zero effective magnetic flux as a function of angular 
momentum $L$.  The results are
shown for a system with $N=27$ particles.  Top panel, electrons;  middle panel 
composite fermions at $\nu=1/2$;  bottom panel, composite fermions of 
$\nu=5/2$.
The quantity $l_0=\sqrt{\hbar/eB}$ is 
the magnetic length, $r_0=(\pi \rho)^{-1/2}$ is the
average interparticle separation, and $\epsilon $ is the
dielectric constant of the background material ($\epsilon =12.8$ for GaAs).
\label{fig2}}
\end{figure}

The energy of $\Psi_L^{CF-pair}$, $E[L]$, is evaluated in a Monte Carlo approach \cite{JK}.
Figure~(\ref{fig2}) shows the energy of the pair as a function of 
$L$ for electrons at $B=0$, and for composite fermions at $\nu=1/2$ and 5/2.
The interaction energy between the two added electrons decreases with increasing $L$ 
at $B=0$, as expected.  Similar behavior is found for composite fermions
at $\nu=1/2$.  However, the opposite behavior is seen at $\nu=5/2$, 
indicating an attractive interaction between composite fermions at $\nu=5/2$. 
The largest binding energy is obtained for the $L=1$ channel.
This qualitative difference between the physics at $\nu=1/2$ and $\nu=5/2$ 
is the principal result of this work.

\begin{figure}
\centerline{\psfig{figure=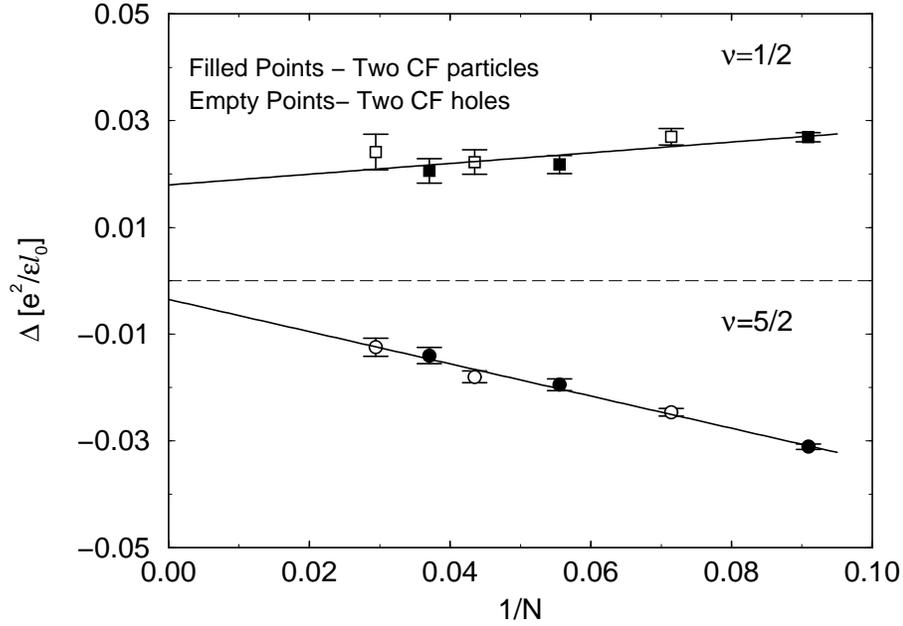,width=14.0cm,angle=-90}}
\caption{The binding energy of the composite fermion (CF)
 pair in the $L=1$ channel 
at $\nu=1/2$ and 5/2 as a function of
$1/N$.  The binding energy is defined as the energy of the pair at $L=1$ relative to
its energy at $L_{max}$.  The filled (empty) symbols are for CF-particles (CF-holes)
on top of the CF-Fermi surface.  As expected from particle-hole symmetry, satisfied by the
wavefunctions considered here to an excellent approximation, the binding energies for the
CF-particles and CF-holes fall on the same line. 
\label{fig3}}
\end{figure}

The estimation of the thermodynamic limit of the binding energy from our finite size
study can be difficult.  When the pair interaction energy is not small 
compared to the inter-shell spacing, many shells would participate in the 
pair wavefunction, and our approximation of restricting the pair 
to the lowest unoccupied shell would break down.  A determination of the appropriate
`parent' wavefunction $\Phi^{el-pair}$ is obviously quite 
complicated in this regime, and consequently, so is obtaining 
$\Psi^{CF-pair}$.  However, past studies \cite {Wu} have shown that 
distinct  excitations in $\Phi$ with the same quantum numbers may produce the same 
excitation $\Psi$, because the Hilbert space at $\nu=1/2$ is greatly 
restricted compared to that at zero magnetic field.
Therefore, as a first step, we proceed without any explicit 
consideration of shell mixing.  We have studied up to $n=6$ filled shells;
both CF-particle pairs and CF-hole pairs are considered 
for up to $n=5$ and only the latter for $n=6$.  Figure~(\ref{fig3}) shows that 
despite a substantial decrease with $N$, the binding energy $\Delta[L=1]=E[L=1]-E[L_{max}]$ 
at $\nu=5/2$ still extrapolates to a non-zero negative value 
of $-0.0035(\pm0.0013)$ $e^2/\epsilon l_0$ as $N\rightarrow \infty$, where 
$l_0=\sqrt{\hbar /eB}$ and $\epsilon$ is the dielectric constant of the background 
material.  After taking account of the transverse width of the electron wavefunction
\cite{Fang,Zhang}, the 
binding energy in the $L=1$ channel is estimated to be $-0.0025(\pm0.0015)$
$e^2/\epsilon l_0$ for densities in the range (0.5-3.0) $\times$10$^{11}$ cm$^{-2}$.
Following the weak-coupling BCS theory,
it is natural to identify $2|\Delta|$ with the energy gap of the FQHE state 
at $\nu=5/2$.  For the parameters of the experiment of Pan {\em et al.} \cite{Pan}, our 
calculated $2|\Delta|$ for $L=1$ is $0.5(\pm0.3)$K, which turns out to be 
on the same order of magnitude as the measured gap  of about 0.1K.  
While this is encouraging, we note that the effect of shell mixing is an important issue, and must be considered to ascertain the quantitative reliability 
of our preliminary estimate.

The appearance of pairing may seem surprising in a model with strong repulsive 
interaction.  However, the Coulomb repulsion is overcome through the formation 
of 
composite fermions, which screens out the Coulomb interaction quite effectively, 
to the extent that a total neglect of the interaction between composite fermions  
is sufficient for many purposes.  
Furthermore, the screening takes place in a topologically rigid manner, independent
of the interaction strength or the Landau level index, through 
the binding of precisely two vortices 
to each electron.  Therefore, it is plausible that  
sometimes an overscreening of the Coulomb
interaction occurs, producing an effectively attractive interaction between composite
fermions.  The reason that there is an attraction at $\nu=5/2$ but not at $\nu=1/2$ is that the matrix  
elements for the Coulomb interaction in the second Landau level are weaker than 
in the lowest LL because of the greater spread of 
the electron wavefunction in the former, especially at short distances.  
For example, $V_{1,1}/V_{0,1}=0.93$.  This slight softening of the inter-electron 
repulsive interaction in the second LL is sufficient to make the inter-CF interaction weakly negative.

There has been earlier work on pairing of composite fermions.
Many years ago, Greiter, Wen, and Wilczek \cite{Greiter} argued for $p$-wave pairing 
of composite fermions at $\nu=1/2$ and 5/2 within a Chern-Simons formulation of
composite fermions.  The Chern-Simons method, however, is quantitatively 
unreliable in this application because of its inadequacy in describing the energetics or the 
short-distance behavior.  Even within
this approach, Bonesteel \cite {Bonesteel} has noted that
a pair breaking term not considered in ref.~\cite{Greiter} may
potentially alter its conclusion.  Greiter {\em et al.} further  
suggested that the paired CF state may be described in terms of a Pfaffian wavefunction written by Moore and Read \cite{Moore}.  Recent exact diagonalization 
\cite{Morf,Rezayi} and variational \cite {PMBJ} studies have provided support for 
the validity of a Pfaffian-like wavefunction at $\nu=5/2$.

The pairing of composite fermions at $\nu=5/2$ has a topological origin, and occurs in spite of strong repulsive interaction 
between electrons.  The repulsion is circumvented because 
the objects that pair up are not electrons but composite fermions.  
We speculate that a fundamental reorganization of the state, for example,  
creation of new quasiparticles, must happen in any system in order for pairing 
to ensue, starting from repulsive interactions alone. This is indeed the case in several  
theoretical models of high-temperature superconductivity, where the pairing 
is believed also to be caused by repulsive interactions.

{\bf Acknowledgements}

This work was supported in part by the National Science Foundation.
We thank the Numerically Intensive Computing Group led by V.
Agarwala, J. Holmes, and J. Nucciarone, at the Penn State University CAC, for
assistance and computing time with the LION-X cluster.

\pagebreak

\pagebreak

\pagebreak

\end{document}